# Some modifications to the SNIP journal impact indicator

Ludo Waltman, Nees Jan van Eck, Thed N. van Leeuwen, and Martijn S. Visser

Centre for Science and Technology Studies, Leiden University, The Netherlands
{waltmanlr, ecknjpvan, leeuwen, visser}@cwts.leidenuniv.nl

The SNIP (source normalized impact per paper) indicator is an indicator of the citation impact of scientific journals. The indicator, introduced by Henk Moed in 2010, is included in Elsevier's Scopus database. The SNIP indicator uses a source normalized approach to correct for differences in citation practices between scientific fields. The strength of this approach is that it does not require a field classification system in which the boundaries of fields are explicitly defined.
In this paper, a number of modifications that will be made to the SNIP indicator are explained, and the advantages of the resulting revised SNIP indicator are pointed out. It is argued that the original SNIP indicator has some counterintuitive properties, and it is shown mathematically that the revised SNIP indicator does not have these properties. Empirically, the differences between the original SNIP indicator and the revised one turn out to be relatively small, although some systematic differences can be observed. Relations with other source normalized indicators proposed in the literature are discussed as well.

## 1. Introduction

The SNIP indicator, where SNIP stands for *source normalized impact per paper*, measures the citation impact of scientific journals using a so-called source normalized approach (Moed, 2010). The idea of the source normalized approach is to correct for differences in citation practices between scientific fields without requiring a field classification system in which the boundaries of fields are explicitly defined. In the source normalized approach, normalization of citation counts for field differences takes place based on the characteristics of the sources from which citations originate. If a journal is cited mainly by publications with long reference lists, this suggests that the journal finds itself in a field with a high citation density (e.g., cell biology), which means that it is reasonable to expect the journal to receive a relatively large number of citations per publication. Conversely, if the publications that cite a journal tend to have short reference lists, the journal would appear to be active in a low citation density field (e.g., mathematics), and only a modest number of citations per publication can be expected for the journal.

The SNIP indicator was introduced by Henk Moed in 2010 (Moed, 2010). The indicator is made available in Elsevier's Scopus database, together with another journal impact indicator, the SCImago Journal Rank (SJR) indicator (González-Pereira, Guerrero-Bote, & Moya-Anegón, 2010; Guerrero-Bote & Moya-Anegón, 2012).[1] The calculation of the SNIP values reported in the Scopus database is done by our institute, the Centre for Science and Technology Studies (CWTS) of Leiden University.

---

[1] The Scopus database can be accessed at www.scopus.com. A subscription is required. The SNIP and SJR indicators are also freely available at www.journalmetrics.com.

In 2012, a number of modifications will be made to the SNIP indicator. Our aim in this paper is to explain these modifications and to discuss the advantages of the revised SNIP indicator over the original one. We also present an empirical analysis in which the original SNIP indicator and the revised one are compared with each other.

In addition to the SNIP indicator, a number of other source normalized indicators have been proposed in the literature. The first proposal was the audience factor of Zitt and Small (2008). An alternative version of this indicator was introduced by Zitt (2010). The SNIP indicator was criticized by Leydesdorff and Opthof (2010), who proposed a different approach to source normalization. Leydesdorff and Opthof referred to their proposed approach as fractional counting of citations. See Leydesdorff and Bornmann (2011), Leydesdorff and Opthof (2011), Leydesdorff, Zhou, and Bornmann (in press), Moed (2011), and Radicchi and Castellano (2012) for further discussion of this idea. A somewhat similar approach was introduced by Glänzel, Schubert, Thijs, and Debackere (2011), under the label of 'a priori normalization'. As we will discuss in this paper (mainly in the technical appendices), the revised SNIP indicator that we introduce has elements in common with a number of these earlier proposals, but the indicator also incorporates some important new ideas.

The organization of the paper is as follows. First, in Section 2, the main elements of the original SNIP indicator are summarized and some properties of the indicator are discussed. Next, in Section 3, the revised SNIP indicator is introduced, with a focus on the differences with the original SNIP indicator. A special issue in the case of the revised SNIP indicator is the selection of citing journals. This issue is considered in Section 4. The empirical analysis is reported in Section 5. Finally, conclusions are drawn in Section 6. A number of technical results are discussed in two appendices.

## 2. Original SNIP indicator

In this section, we provide an overview of the main elements of the original SNIP indicator. We also discuss some properties of the indicator. We refer to Moed (2010) for a more comprehensive treatment of the original SNIP indicator.

The original SNIP indicator is defined as the ratio of a journal's *raw impact per paper* (RIP) and a journal's *relative database citation potential* (RDCP), that is,

$$\text{SNIP} = \frac{\text{RIP}}{\text{RDCP}}. \qquad (1)$$

For a given year of analysis, the RIP value of a journal equals the average number of times the journal's publications in the three preceding years were cited in the year of analysis. For instance, if 100 publications appeared in a journal in the period 2008–2010 and if these publications were cited 200 times in 2011, the RIP value of the journal for 2011 equals 200 / 100 = 2. In the calculation of RIP values, citing and cited publications are included only if they have the Scopus document type *article*, *conference paper*, or *review*. We note that the RIP indicator is similar to the well-known journal impact factor, although the RIP indicator uses three instead of two years of cited publications and includes only citations to publications of specific document types.

The RIP indicator reflects the average citation impact of the publications of a journal, without correcting for differences in citation practices between scientific fields. Because there is no normalization for field differences, RIP values should not be compared across fields. By dividing a journal's RIP value by its RDCP value, the SNIP indicator aims to provide a measure of citation impact that allows for meaningful between-field comparisons. The RDCP value of a journal is given by

$$\text{RDCP} = \frac{\text{DCP}}{\text{median(DCP)}}, \qquad (2)$$

where DCP denotes a journal's *database citation potential* and median(DCP) denotes the median DCP value of all journals in the database.[2] It follows from (2) that the median RDCP value of all journals in the database equals one. As a consequence, half of the journals in the database have a SNIP value that is higher than their RIP value and half of the journals in the database have a SNIP value that is lower than their RIP value. We note that the division by median(DCP) in (2) has the same effect for all journals. It influences journals' SNIP values in an absolute sense, but not in a relative one. Hence, within a given year of analysis, the division by median(DCP) has no effect on the way in which journals compare with each other.

Since the step from a journal's DCP value to its RDCP value boils down to a division by a fixed value, what is really important is to understand the calculation of the DCP value of a journal. This calculation starts by delineating a journal's subject field. The subject field of a journal is defined as the set of all publications in the year of analysis with at least one reference to the journal.[3] The DCP value of a journal equals the average number of references in the publications in the subject field of the journal, counting only references to publications that appeared in the three preceding years in journals covered by the database. Like in the calculation of the RIP value of a journal, the calculation of a journal's DCP value takes into account only citing and cited publications of the Scopus document types *article*, *conference paper*, and *review*. Mathematically, the DCP value of a journal can be expressed as

$$\text{DCP} = \frac{r_1 + r_2 + \ldots + r_n}{n}, \qquad (3)$$

where $n$ denotes the number of publications in the subject field of the journal and $r_i$ denotes the number of references in the $i$th publication to publications that appeared in the three preceding years in journals covered by the database. In the rest of this paper, we will refer to the references included in $r_i$ as active references. This follows the terminology introduced by Zitt and Small (2008).

In what way does the division in (1) of a journal's RIP value by its RDCP value yield a normalization for field differences? Somewhat informally, this question can be answered as follows. Consider two scientific fields, field X and field Y. Suppose that publications in field X on average have 12 active references. Publications in field Y on average have 3 active references. Assuming that publications mostly refer to other publications in their own field and that the number of publications in a field is fairly stable over time, it can be seen that the average RIP value of journals in field X must be close to 4, while in field Y the average RIP value of journals must be about 1.[4] If it is further assumed that the subject field of a journal consists of a random selection of publications from the field to which the journal belongs, it follows that journals in field X must have DCP values around 12, while the DCP values of journals in field Y must be close to 3. Taking into account that the effect of the division by median(DCP) in (2) is the same for both fields, it is clear that the difference in RIP values between fields X and Y will be more or less canceled out by the difference in DCP values between the two fields. The SNIP values of the journals in the two fields will therefore be in approximately the same range, and in this sense a correction for field differences will have been made.

---

[2] The notion of citation potential originates from Garfield (1979). See Moed (2010) for more details.
[3] In the definition of the subject field of a journal, only references going back no more than eight years are considered.
[4] For instance, suppose that in the year of analysis 1,000 publications appeared in field X. Overall, these publications have 12,000 active references. These references point to publications that appeared

Although the above reasoning shows that under appropriate assumptions the normalization mechanism of the original SNIP indicator works in a satisfactory way, the mechanism also has some properties that we consider counterintuitive. We will now discuss two of these properties.

The first property is about the effect of receiving an additional citation on the SNIP value of a journal. It turns out that in some cases receiving an additional citation may have a negative effect on a journal's SNIP value. This may happen if the citing publication has a lot of active references. To see this, suppose that in a given year of analysis 80 citations were given to the 10 publications that appeared in a particular journal in the three preceding years. On average, the 80 citing publications have 4 active references. The RIP value of the journal then equals $80 / 10 = 8$, while the DCP value equals 4. Suppose that the median DCP value of all journals in the database equals 2. According to (2), our journal of interest then has an RDCP value of $4 / 2 = 2$. Based on (1), this means that the SNIP value of the journal equals $8 / 2 = 4$. Now assume that an additional citation is given to the journal. The citing publication has a long reference list, with 100 active references. What does this mean for the SNIP value of the journal? The journal's RIP value increases from 8 to $81 / 10 = 8.10$. Its DCP value increases as well, from 4 to $(80 \times 4 + 100) / 81 \approx 5.19$, yielding an RDCP value of $5.19 / 2 \approx 2.59$. Hence, in the end, the SNIP value of our journal of interest goes down from 4 to $8.10 / 2.59 \approx 3.12$. In other words, according to the original SNIP indicator, the additional citation received by the journal implies a decrease in the normalized citation impact of the journal. In our view, this result is counterintuitive.

Another property of the original SNIP indicator that we consider counterintuitive relates to the notion of consistency discussed earlier in the context of other field-normalized indicators (Waltman, Van Eck, Van Leeuwen, Visser, & Van Raan, 2011) as well as in the context of the *h*-index (Waltman & Van Eck, 2012a). The original SNIP indicator turns out not to satisfy certain consistency requirements. To illustrate this, we consider the situation in which two journals are merged. It seems reasonable to expect that the merger of two journals should lead to a new journal with a SNIP value that is in between the SNIP values of the two original journals. Using the original SNIP indicator, however, this need not always be the case. For instance, suppose that two journals, journal X and journal Y, are merged. Statistics for these journals are reported in Table 1. As can be seen in the table, both journals have the same number of publications. The RIP and DCP values of journal Y are twice as high as those of journal X. However, in the SNIP calculation this cancels out, and both journals therefore have the same SNIP value of 6. We would expect the new journal

---

in field X in the three preceding years. There are 3,000 such publications. This then implies that on average each publication receives $12,000 / 3,000 = 4$ citations. The average RIP value of journals in field X therefore equals 4. Using a similar reasoning, an average RIP value of 1 is obtained for journals in field Y.

resulting from the merger of journals X and Y to also have a SNIP value of 6, but Table 1 shows that this is not the case. The RIP value of the new journal, referred to as journal XY in Table 1, equals (120 + 240) / (10 + 10) = 18, which is simply the average of the RIP values of journals X and Y. However, the DCP value of journal XY does not equal the average of the DCP values of journals X and Y. There are 120 + 240 = 360 publications citing journal XY,[5] and overall these publications have 120 × 6 + 240 × 12 = 3,600 active references. This results in a DCP value of 3,600 / 360 = 10, which is closer to the DCP value of journal Y than to the DCP value of journal X. In the end, a SNIP value of 5.40 is obtained for journal XY. This is lower than the SNIP values of both journal X and journal Y. Hence, according to the original SNIP indicator, merging journals X and Y yields a new journal that has a lower normalized citation impact than each of the original journals. This is something we consider counterintuitive.

Table 1. Example of the calculation of the original SNIP indicator in a situation in which two journals are merged.

|  | Journal X | Journal Y | Journal XY |
| --- | --- | --- | --- |
| No. pub. | 10 | 10 |  |
| No. cit. | 120 | 240 |  |
| No. active ref. per citing pub. | 6 | 12 |  |
| RIP | 12 | 24 | 18 |
| DCP | 6 | 12 | 10 |
| median(DCP) | 3 | 3 | 3 |
| RDCP | 2 | 4 | 3.33 |
| SNIP | 6 | 6 | 5.40 |

We have now discussed two properties of the original SNIP indicator that we regard as unsatisfactory. The revised SNIP indicator, introduced in the next section, is similar to the original SNIP indicator but includes a number of modifications. Because of these modifications, the revised SNIP indicator does not have the unsatisfactory properties of the original SNIP indicator.

## 3. Revised SNIP indicator

Compared with the original SNIP indicator, the three most significant modifications made in the revised SNIP indicator can be summarized as follows:

- DCP values are calculated as harmonic rather than arithmetic averages.
- The calculation of DCP values takes into account not only the number of active references in citing publications but also the proportion of publications with at least one active reference in citing journals.
- The distinction between DCP and RDCP values is abandoned.[6]

---

[5] It is assumed here that each citation to journals X and Y originates from a different citing publication. Also, for simplicity, we ignore the possibility that there may be additional publications in the subject fields of journals X and Y (i.e., publications citing older publications in these journals).

[6] In the case of the original SNIP indicator, the distinction between DCP and RDCP values is made in order to ensure that half of the journals in the database have a SNIP value that is higher than their RIP value and half of the journals in the database have a SNIP value that is lower than their RIP value. In the case of the revised SNIP indicator, the distinction between DCP and RDCP values is abandoned because SNIP values are standardized differently. As discussed below, in the case of the revised SNIP indicator, DCP values are calculated in such a way that the average SNIP value of all journals in the database is close to one.

Notice that all modifications made in revised SNIP indicator relate to the normalization mechanism that is used. The modifications therefore affect only the denominator in (1). The RIP indicator in the numerator is not affected.

The revised SNIP indicator is defined as the ratio of a journal's RIP value and its DCP value. In mathematical terms,

$$\text{SNIP} = \frac{\text{RIP}}{\text{DCP}}. \tag{4}$$

The RIP value of a journal is calculated in exactly the same way as in the case of the original SNIP indicator. Hence, for a given year of analysis, the RIP value of a journal equals the average number of times the journal's publications in the three preceding years were cited in the year of analysis.

In the case of the original SNIP indicator, the DCP value of a journal equals the average number of active references in the publications belonging to the journal's subject field, where the average is calculated as an ordinary arithmetic average, as indicated in (3). In some situations, it makes more sense to use an alternative type of average instead of the commonly used arithmetic average.[7] As we will explain below, in the calculation of a journal's DCP value, we consider the use of a harmonic average more sensible than the use of an arithmetic average. In the case of the revised SNIP indicator, we therefore calculate a journal's DCP value as

$$\text{DCP} = \frac{1}{3} \times \frac{n}{\frac{1}{p_1 r_1} + \frac{1}{p_2 r_2} + \ldots + \frac{1}{p_n r_n}}, \tag{5}$$

where $n$ and $r_i$ are defined in the same way as in the previous section, that is, $n$ denotes the number of publications in the subject field of the journal and $r_i$ denotes the number of active references in the $i$th publication in the journal's subject field.

Compared with (3), what is new in (5) is $p_i$. The definition of $p_i$ is not entirely trivial. To determine $p_i$, we take the following three steps: (1) We first select the $i$th publication in the subject field of a journal, (2) we then look at all publications that appeared in the same journal and in the same year as the selected publication, and (3) finally we calculate $p_i$ as the proportion of these publications that have at least one active reference. In fields with a high citation density (e.g., cell biology), $p_i$ will typically be close to one, since in most journals in these fields all or almost all publications will have active references. On the other hand, in fields with a low citation density (e.g., mathematics), a considerable share of the publications may have no active references, and $p_i$ may therefore be significantly below one. The justification for including $p_i$ in (5) is somewhat technical and is provided in the appendices, along with some other technical results. The general idea is that without including $p_i$ source normalization mechanisms may fail to completely correct for differences between low citation density and high citation density fields (for some empirical evidence of this problem, see Glänzel et al., 2011; Leydesdorff & Bornmann, 2011). Including $p_i$ should solve this problem.

---

[7] A typical example is the calculation of the average speed of a car. Suppose that a car travels a distance of 100 km at a speed of 50 km per hour and another distance of 100 km at a speed of 200 km per hour. The average speed of the car then does not equal the arithmetic average of (50 + 200) / 2 = 125 km per hour. To see this, notice that the total travel time is 100 / 50 + 100 / 200 = 2.5 hour, resulting in an average speed of (100 + 100) / 2.5 = 80 km per hour. To calculate the average speed of the car, one needs to use the harmonic instead of the arithmetic average, which yields the correct result of 2 / (1 / 50 + 1 / 200) = 80 km per hour.

A number of additional comments need to be made on the DCP calculation in (5):

- Like in the DCP calculation in the original SNIP indicator, only publications of the Scopus document types *article*, *conference paper*, and *review* are considered as citing and cited publications.
- The subject field of a journal consists of all publications in the year of analysis that refer to a publication in the journal in the three preceding years.[8] In the case of the original SNIP indicator, there can be no duplicate publications in the subject field of a journal. This is different in the case of the revised SNIP indicator. The revised SNIP indicator allows for multiple occurrences of the same publication in a journal's subject field. For instance, if a publication refers to five publications in a particular journal in the three preceding years, the publication is counted five times in the journal's subject field.
- The DCP calculation in (5) includes a multiplication by 1/3. This multiplication ensures that the average SNIP value of all journals in the database is close to one. We refer to Appendix B for more details.

What remains to be discussed is why in (5) a journal's DCP value is calculated as a harmonic average. Why does it make more sense to use a harmonic average instead of an arithmetic one? This can be illustrated by returning to the example of the merger of two journals, which we introduced at the end of the previous section. Statistics for the two journals to be merged, referred to as journal X and journal Y, as well as for the merged journal, referred to as journal XY, are reported in Table 2. These statistics are based on the revised SNIP indicator, which means that a journal's DCP value is calculated as a harmonic average using (5).

Table 2. Example of the calculation of the revised SNIP indicator in a situation in which two journals are merged.

|  | Journal X | Journal Y | Journal XY |
|---|---|---|---|
| No. pub. | 10 | 10 |  |
| No. cit. | 120 | 240 |  |
| No. active ref. per citing pub. | 6 | 12 |  |
| RIP | 12 | 24 | 18 |
| DCP | 2 | 4 | 3 |
| SNIP | 6 | 6 | 6 |

Comparing Table 2 with Table 1, a crucial difference can be observed. In Table 1, in which statistics based on the original SNIP indicator are presented, we observe the counterintuitive result that the SNIP value of journal XY is lower than the SNIP values of both journal X and journal Y. In Table 2, on the other hand, the SNIP value of journal XY equals the SNIP values of journals X and Y. In our view, this is the result that one would expect intuitively. As can be seen, the difference between Tables 1 and 2 is due to the way in which the DCP value of journal XY is calculated. In Table 2, the DCP value of journal XY calculated using (5) equals $1/3 \times (120 + 240) / (120 \times (1/6) + 240 \times (1/12)) = 3$.[9] In other words, the DCP value of journal XY is exactly halfway in between the DCP values of journals X and Y. This is how we believe it should be, since in the end this yields the desired result that journal XY has the same SNIP value as journals X and Y. In Table 1, on the other hand, the use of an arithmetic average in (3) causes the DCP value of journal XY to be closer to journal Y's DCP value than to journal X's, and this in turn leads to the unsatisfactory result that the SNIP value of journal XY is lower than the SNIP values of both journal X and journal Y. In summary, our example of the merger of two journals shows a situation in which the use of a harmonic average to calculate a journal's DCP value yields a more intuitively reasonable result than the use of an arithmetic average. This provides some support for the use of a harmonic instead of an arithmetic average in the calculation of journals' DCP values.

---

[8] In the case of the original SNIP indicator, the subject field of a journal is defined based on citations going back at most eight years. In the case of the revised SNIP indicator, a three-year citation window is used, which is consistent with the citation window used in the calculation of RIP values.

[9] For simplicity, we assume that $p_i = 1$ for each citing publication.

To provide further support for the revised SNIP indicator, a more formal mathematical analysis is needed. Such an analysis is presented in Appendices A and B. In Appendix A, we look at the revised SNIP indicator from an alternative perspective by rewriting (4) and (5) given above. The alternative perspective that we take allows us to compare the revised SNIP indicator with some other source normalized indicators proposed in the literature (Glänzel et al., 2011; Leydesdorff & Bornmann, 2011; Leydesdorff & Opthof, 2010; Zitt & Small, 2008). It also makes it possible to show mathematically that the revised SNIP indicator does not have the two counterintuitive properties of the original SNIP indicator that we discussed at the end of the previous section. In Appendix B, we investigate the way in which the revised SNIP indicator corrects for differences in citation practices between scientific fields. We prove that under certain assumptions field differences are indeed properly corrected for.

## 4. Selection of citing journals

Source normalization provides an elegant mechanism for correcting for field differences without requiring a field classification system in which the boundaries of fields are explicitly defined. However, source normalization also has an important weakness. As already indicated by Zitt and Small (2008), it has difficulties to properly handle citing journals with special referencing behavior. Examples include trade journals, scientific magazines, and scientific journals with a strong national focus. These journals tend to have a relatively small average number of active references per publication, much smaller than what is common for 'regular' scientific journals. Because of the small numbers of active references, citations from these journals tend to significantly decrease the DCP value of the cited journal and, consequently, to significantly increase the cited journal's SNIP value. In this way, we may end up in the counterintuitive situation that the SNIP value of a journal benefits more from a citation from a trade journal or some obscure national scientific journal than from a citation from a 'regular' international scientific journal.

The strong sensitivity of SNIP values to citations from special types of citing journals is something we consider undesirable. In the calculation of the revised SNIP indicator, journals with very small numbers of active references are therefore excluded as citing journals. This means that citations given by these journals are not taken into account in the calculation of the revised SNIP indicator. We note that journals excluded as citing journals do have a SNIP value, just like all other journals.[10]

---

[10] However, the SNIP values of journals that are excluded as citing journals do not benefit from journal self citations. This may cause a disadvantage for these journals.

To determine which journals are included as citing journals and which are not, we first correct for journal title changes. If the title of a journal has changed, Scopus treats the journal as two different journals, one with the old title and one with the new title. We merge such journals into a single journal. After correcting for journal title changes, journals are excluded as citing journals in the following three steps:

1. *Exclude trade journals*. Each journal in Scopus has a classification. In the first step, journals classified as trade journal are excluded as citing journals.
2. *Exclude journals that did not publish continuously during four consecutive years*. For a given year of analysis, a journal is excluded as a citing journal if it does not have publications both in the year of analysis and in each of the three preceding years. Hence, a journal can be included as a citing journal only if it has publications during four consecutive years. This restriction aims to create an appropriate balance between the number of citing publications and the number of cited publications in the calculation of the revised SNIP indicator.[11]
3. *Exclude journals for which less than 20% of the publications in the year of analysis have at least one active reference*. A journal is included as a citing journal if in the year of analysis at least 20% of its publications have at least one active reference. An active reference is defined as a reference to a publication that appeared in one of the three preceding years in a journal included as a citing journal in the present year.[12]

The effect of the above rules for excluding journals as citing journals will be discussed in the empirical analysis presented in the next section.

## 5. Empirical analysis

In this section, we report an empirical analysis of the revised SNIP indicator and we empirically compare the revised SNIP indicator with the original one. The most recent year for which values of the original SNIP indicator are available is 2010, and we therefore use this year as our year of analysis. This means that SNIP values are calculated based on citations from 2010 to 2007, 2008, and 2009.

### 5.1. Citing journals

The selection of citing journals follows the three steps discussed in Section 4. Of the 19,816 journals with at least one publication[13] in 2010, 262 were excluded as citing journals because they have a trade journal classification in Scopus. Another 5,568 journals were excluded because they did not have publications in each year in the period 2007–2010. Among the remaining 13,986 journals, there were 832 for which the proportion of publications that have at least one active reference was below 20%. These journals were excluded as well, which means that in the end there were 13,154 citing journals left.

---

[11] The importance of having an appropriate balance between the number of citing publications and the number of cited publications in the calculation of the revised SNIP indicator is discussed in Appendix B.

[12] Notice the recursiveness in these definitions: Whether a journal is included as a citing journal depends on whether it has sufficient active references, and whether a reference counts as active depends on whether it points to a journal included as a citing journal. This recursiveness is dealt with by iteratively applying the definitions of a citing journal and an active reference until a stable set of citing journals is obtained. We further note that the threshold of 20% publications with at least one active reference was chosen because it seems to give a reasonable trade-off between on the one hand excluding the most problematic journals and on the other hand including as many journals as possible.

[13] We use the term 'journal' to refer to all sources included in Scopus, not only journals but also conference proceedings and book series. When referring to publications, we mean publications of the Scopus document types *article*, *conference paper*, and *review*. In addition, we exclude publications that have no references. In many cases, these publications actually do have references, but data on their references is missing in Scopus. Also, some publications appear two times in Scopus, one time with references and one time without.

**5.2. Selected results**

Values of the revised SNIP indicator were calculated for all 22,434 journals with at least one publication in the period 2007–2009 (excluding trade journals). Only citations from the 13,154 journals selected as citing journals were taken into account. The average of the SNIP values of the citing journals, where each journal's SNIP value is weighted by the number of publications of the journal in the period 2007–2009, equals 1.03. This is slightly above the theoretically expected value of one (see Appendix B). The difference is due to the fact that the publication output of the citing journals in 2010 is 3% larger than the average publication output of the same journals in the years 2007, 2008, and 2009. Looking at all 22,434 journals with a SNIP value instead of only the 13,154 citing journals, an average SNIP value of 0.90 is obtained.

Table 3 lists the top 30 journals with the highest SNIP value in 2010. Only journals with at least 100 publications in the period 2007–2009 are shown. As is to be expected, many review journals are listed in Table 3. Ignoring these journals, the table displays a broad range of journals from many different scientific fields. In addition to the multidisciplinary journals *Nature* and *Science*, we observe journals from the medical sciences, the life sciences, and the natural sciences, but also computer science and engineering journals as well as an economics journal. The diversity of journals in the top of the SNIP ranking can be seen as an indication that the revised SNIP indicator successfully corrects for differences in citation practices between fields.

Table 3. Top 30 journals with the highest value for the revised SNIP indicator (excluding journals with fewer than 100 publications).

| Journal | SNIP | Journal | SNIP |
| --- | --- | --- | --- |
| Acta Crystallographica Section A | 35.47 | The Lancet Neurology | 6.76 |
| Reviews of Modern Physics | 27.75 | Nature Genetics | 6.49 |
| New England Journal of Medicine | 13.35 | IEEE Journal on Selected Areas in Communications | 6.36 |
| Physics Reports | 11.81 | Nature Physics | 6.31 |
| Chemical Reviews | 11.12 | Chemical Society Reviews | 6.26 |
| Physiological Reviews | 10.58 | Psychological Bulletin | 6.25 |
| Progress in Polymer Science | 10.03 | Quarterly Journal of Economics | 6.18 |
| JAMA | 9.37 | Cell | 6.10 |
| Nature Materials | 8.52 | Nature Biotechnology | 6.09 |
| Nature | 8.51 | Annals of Internal Medicine | 6.07 |
| Clinical Microbiology Reviews | 8.43 | Academy of Management Review | 5.99 |
| Science | 7.89 | IEEE Trans. on Software Engineering | 5.95 |
| Nature Nanotechnology | 7.26 | Proceedings of the IEEE | 5.94 |
| Nature Photonics | 7.19 | IEEE Signal Processing Magazine | 5.92 |
| IEEE Trans. on Pattern Analysis and Machine Intelligence | 6.91 | International Journal of Computer Vision | 5.74 |

Figure 1 shows the relation between the RIP indicator and the revised SNIP indicator for all 10,347 journals with at least 100 publications in the period 2007–2009. On average, the RIP value of a journal is 1.85 times as high as the SNIP value. This is indicated by the diagonal line in Figure 1, which shows the average relation between the RIP indicator and the revised SNIP indicator. The journals located far below the line turn out to be mainly from the life sciences, but also from the medical sciences and from chemistry. *Nature* and *Science* are also among these journals. The journals located at the largest distance above the line turn out to be from computer science, economics, and mathematics, which are well-known low citation density fields. Some physics journals are located significantly above the line as well.

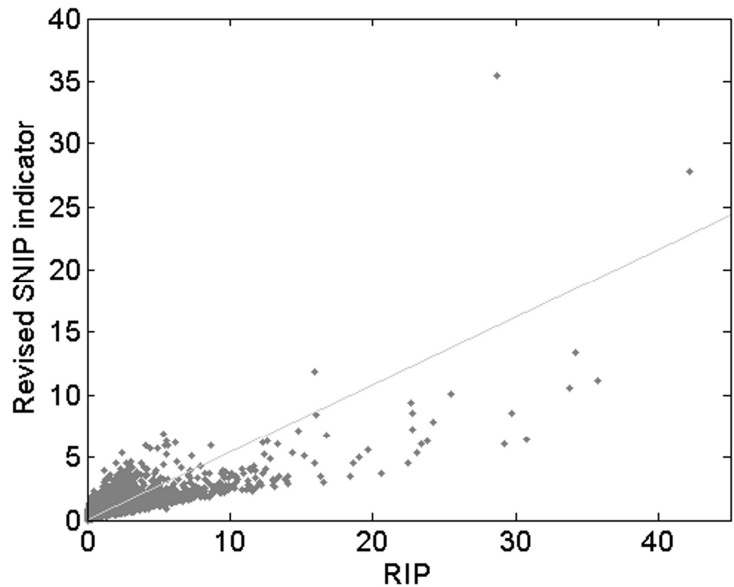

Figure 1. RIP and revised SNIP values of all journals with at least 100 publications. The Pearson correlation coefficient equals 0.79.

**5.3. Comparison with the original SNIP indicator**

For our year of analysis (i.e., 2010), there are 22,403 journals for which values are available both for the original SNIP indicator and for the revised one. The average of the original SNIP values of these journals, where each journal's SNIP value is weighted by the publication output of the journal, equals 1.13. This is about 26% higher than the average of the revised SNIP values of the journals, which equals 0.90. Hence, as a consequence of the differences in calculation discussed in Section 3, the revised SNIP indicator on average yields lower values than the original SNIP indicator.

Figure 2 shows the relation between the original SNIP indicator and the revised one for the 10,331 journals with at least 100 publications in the period 2007–2009. The diagonal line in the figure indicates the average relation between the two indicators based on the above mentioned 26% difference between original and revised SNIP values. As can be seen in the figure, the relation between the original SNIP indicator and the revised one is quite strong. This is confirmed by the Pearson correlation coefficient, which equals 0.93.

Table 4 lists the top 30 journals with the highest value for the original SNIP indicator, where only journals with at least 100 publications are included. Comparing Table 4 with Table 3, the two tables turn out to have 21 journals in common. Looking at the nine journals that are listed in Table 4 but not in Table 3, it is remarkable to see that these journals all have a computer science or engineering focus. Apparently, computer science and engineering journals tend to do better in the case of the original SNIP indicator than in the case of the revised SNIP indicator. In fact, looking at Table 4, computer science and engineering journals may seem to be somewhat overrepresented among the journals ranked highest based on the original SNIP indicator.

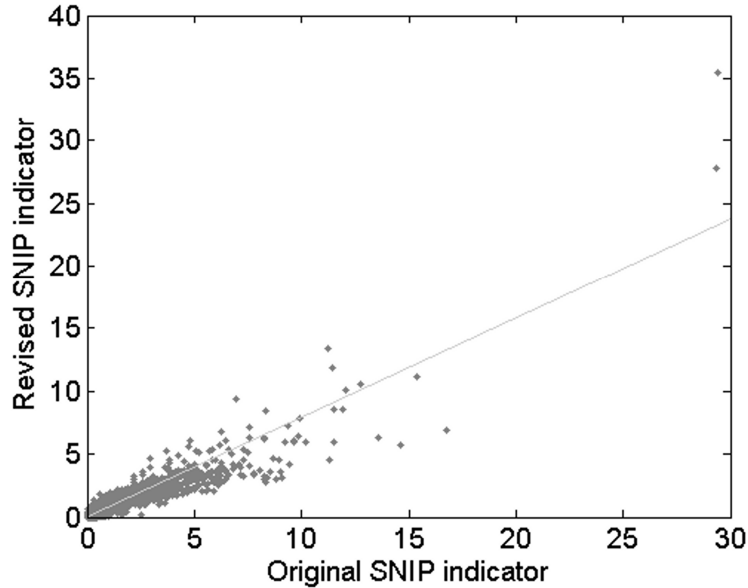

Figure 2. Original and revised SNIP values of all journals with at least 100 publications. The Pearson correlation coefficient equals 0.93.

Table 4. Top 30 journals with the highest value for the original SNIP indicator (excluding journals with fewer than 100 publications).

| Journal | SNIP | Journal | SNIP |
|---|---|---|---|
| Acta Crystallographica Section A | 29.39 | Science | 9.90 |
| Reviews of Modern Physics | 29.32 | Nature Genetics | 9.88 |
| IEEE Trans. on Pattern Analysis and Machine Intelligence | 16.73 | Proceedings of the IEEE | 9.64 |
| Chemical Reviews | 15.37 | Cell | 9.61 |
| International Journal of Computer Vision | 14.64 | IEEE Communications Magazine | 9.40 |
| IEEE Journal on Selected Areas in Communications | 13.59 | Nature Nanotechnology | 9.36 |
| Physiological Reviews | 12.76 | Academy of Management Review | 9.22 |
| Progress in Polymer Science | 12.08 | IEEE Trans. on Industrial Electronics | 9.11 |
| Nature | 11.90 | ACM Trans. on Graphics | 9.04 |
| IEEE Signal Processing Magazine | 11.51 | IEEE/ACM Trans. on Networking | 8.94 |
| Nature Materials | 11.49 | IEEE Trans. on Energy Conversion | 8.81 |
| Physics Reports | 11.45 | IEEE Trans. on Mobile Computing | 8.65 |
| IEEE Trans. on Evolutionary Computation | 11.30 | IEEE Trans. on Image Processing | 8.53 |
| New England Journal of Medicine | 11.25 | IEEE Trans. on Circuits and Systems for Video Technology | 8.39 |
| IEEE Trans. on Software Engineering | 10.22 | Clinical Microbiology Reviews | 8.35 |

For each journal, the difference between the revised SNIP indicator and the original one may be calculated as follows:

$$\text{difference} = \text{SNIP}_{\text{revised}} - \frac{1}{1.26}\text{SNIP}_{\text{original}}. \tag{6}$$

The factor 1/1.26 in (6) follows from the above mentioned observation that for our year of analysis the original SNIP values of journals are on average 26% higher than the revised SNIP values. Table 5 lists both the top 10 journals with the largest positive difference and the top 10 journals with the largest negative difference calculated using (6). Only journals with at least 100 publications are taken into account in the table. In line with our observations based on Tables 3 and 4, negative differences are found for computer science and engineering journals, indicating that the citation impact of these journals is assessed less favorably by the revised SNIP indicator than by the original one. Positive differences are observed for journals in the medical sciences, the life sciences, and the natural sciences.

Table 5. Top 10 journals with the largest positive (left column) or the largest negative (right column) difference between the revised SNIP indicator and the original SNIP indicator (excluding journals with fewer than 100 publications). The difference between the indicators has been calculated using (6).

| Journal | Diff. | Journal | Diff. |
|---|---|---|---|
| Acta Crystallographica Section A | 12.07 | IEEE Trans. on Pattern Analysis and Machine Intelligence | -6.41 |
| Reviews of Modern Physics | 4.40 | International Journal of Computer Vision | -5.92 |
| New England Journal of Medicine | 4.39 | IEEE Journal on Selected Areas in Communications | -4.47 |
| JAMA | 3.83 | IEEE Trans. on Evolutionary Computation | -4.46 |
| Physics Reports | 2.69 | IEEE Trans. on Industrial Electronics | -4.15 |
| Orphanet Journal of Rare Diseases | 2.30 | IEEE Trans. on Energy Conversion | -4.01 |
| Annals of Internal Medicine | 2.25 | IEEE Trans. on Power Systems | -3.82 |
| Acta Crystallographica Section D | 2.23 | ACM Trans. on Graphics | -3.65 |
| Nature Reviews Drug Discovery | 1.87 | IEEE Trans. on Power Electronics | -3.60 |
| Clinical Microbiology Reviews | 1.78 | IEEE Trans. on Circuits and Systems for Video Technology | -3.50 |

## 6. Conclusions

Traditionally, citation-based bibliometric indicators have corrected for differences in citation practices between scientific fields by requiring the boundaries of fields to be explicitly defined in a field classification system. In most cases, fields are delineated at the level of journals in these classification systems. Source normalized indicators, such as the SNIP indicator, have important advantages over traditional normalization approaches based on journal-level classification systems. These traditional approaches for instance have problems with multidisciplinary journals and with fields consisting of subfields with different citation practices. Source normalized indicators do not rely on classification systems and therefore avoid these problems.

In this paper, we have discussed a number of modifications that will be made to the SNIP indicator. We have argued that the original SNIP indicator has some properties that may be considered counterintuitive. For instance, it is possible, at least in theoretical examples, that receiving an additional citation causes a journal's SNIP value to decrease rather than increase. Also, if two journals are merged, the resulting journal may have a lower SNIP value than each of the original journals. The revised SNIP indicator that we have introduced in this paper is defined in such a way that these types of counterintuitive behavior are not possible. We refer to the technical appendices for a more detailed analysis of a number of attractive properties of the revised SNIP indicator.

Our empirical analysis has shown that from an empirical point of view the differences between the original SNIP indicator and the revised one are relatively small. Nevertheless, some systematic differences between the results produced by the two indicators have been revealed. In particular, computer science and engineering journals, which in the case of the original SNIP indicator may seem to be somewhat overrepresented among the top ranked journals, tend to go down in ranking when moving to the revised SNIP indicator.

Finally, let us emphasize that the revised SNIP indicator also has a number of limitations. First of all, the revised SNIP indicator, like most journal impact indicators, is defined as an average of the citation scores of a journal's publications. Given the well-known skewness of citation distributions, it is important to keep in mind the strong sensitivity of average-based indicators to 'outliers', that is, to publications with a very large number of citations. One such publication may be sufficient to cause a large increase in the SNIP value of a journal. This is in fact what has happened with *Acta Crystallographica Section A*, the journal with the highest revised SNIP value in our analysis (see Table 3). Future research may focus on developing alternative types of source normalized indicators that are less sensitive to very highly cited publications.

Two other limitations of the revised SNIP indicator seem to be more or less intrinsic to the idea of source normalization (Zitt & Small, 2008). First, source normalization does not correct for differences between fields in the growth rate of the literature. And second, source normalization does not correct for unidirectional citation flows from one field to another (e.g., from an applied field to a more basic field). These issues are investigated in more detail in Appendix B.

## Appendix A: Some properties of the revised SNIP indicator

In this appendix, we reformulate the mathematical equations underlying the revised SNIP indicator. In this way, an alternative perspective on the revised SNIP indicator is obtained. Based on this alternative perspective, we compare the revised SNIP indicator with some other source normalized indicators proposed in the literature. We also study some properties of the revised SNIP indicator.

The RIP value of a journal can be written as

$$\text{RIP} = \frac{n}{m}, \quad (A1)$$

where $n$ denotes the number of citations received by the journal and $m$ denotes the number of publications of the journal. Combining (A1) with (4) and (5) from Section 3, the revised SNIP indicator can be rewritten as

$$\text{SNIP} = \frac{3}{m} \sum_{i=1}^{n} \frac{1}{p_i r_i}. \quad (A2)$$

It follows from (A2) that the revised SNIP indicator can be interpreted as a journal's average number of citations per publication, where each citation is weighted inversely proportional to both the number of active references in the citing publication and the proportion of publications with at least one active reference in the citing journal.

Looking at the revised SNIP indicator from the perspective of (A2) sheds some light on the way in which the indicator relates to some other source normalized indicators proposed in the literature. The audience factor introduced by Zitt and Small (2008) is similar to the revised SNIP indicator but weights citations inversely proportional to the citing journal's average number of active references per publication. The fractional counting approach proposed by Leydesdorff and Opthof (2010) and Leydesdorff and Bornmann (2011) and the a priori normalization approach studied by Glänzel et al. (2011) are also similar to the revised SNIP indicator, but these approaches differ from (A2) because they do not include something analogous to $p_i$. (The importance of including $p_i$ in the normalization will become clear in Appendix B.) In the case of the fractional counting approach, another difference with the revised SNIP indicator is that normalization takes place based on the total number of references in a citing publication, not only based on the number of active references.

Essentially, the main difference between the various source normalized indicators discussed above can be summarized as follows (for an empirical comparison, see Waltman & Van Eck, 2012b). In the case of the audience factor of Zitt and Small (2008), normalization is done based on the characteristics of citing journals. A potential weakness of this approach is that it may not deal very well with citing journals that cover multiple fields with different citation practices. In the approaches of Leydesdorff and Opthof (2010), Leydesdorff and Bornmann (2011), and Glänzel et al. (2011), normalization is done based on the characteristics of citing publications rather than citing journals. This may be expected to be more accurate, but accuracy may be diminished due to differences between journals in the proportion of publications without active references. The revised SNIP indicator aims to overcome the weaknesses of the above indicators by performing a normalization in which both the characteristics of citing publications (i.e., the number of active references) and the characteristics of citing journals (i.e., the proportion of publications with at least one active reference) are taken into account. Under certain assumptions, this normalization can be proven to properly correct for field differences (see Appendix B).

In Section 2, we discussed two properties of the original SNIP indicator that we consider counterintuitive. One property is that receiving an additional citation may in some situations lead to a decrease in the SNIP value of a journal. The other property is that in the case of a merger of two journals the new journal may have a SNIP value that is not in between the SNIP values of the two original journals. Based on (A2), it can be shown that the revised SNIP indicator does not have the counterintuitive properties of the original SNIP indicator.

First, it is immediately obvious from (A2) that in the case of the revised SNIP indicator receiving an additional citation always leads to an increase in the SNIP value of the cited journal. A citation from a publication with $r$ active references in a journal with a proportion $p$ of publications that have at least one active reference increases the SNIP value of the cited journal by $3 / mpr$. Hence, the larger the number of active references in the citing publication and the larger the proportion of publications with at least one active reference in the citing journal, the smaller the increase in the SNIP value of the cited journal. However, unlike in the case of the original SNIP indicator, the increase is always positive.

Eq. (A2) also implies that in the case of a merger of two journals the new journal always has a SNIP value that is in between the SNIP values of the two original journals. To see this, suppose that the SNIP values of journals X and Y are given by

$$\text{SNIP}^{\text{X}} = \frac{3}{m^{\text{X}}} \sum_{i=1}^{n^{\text{X}}} \frac{1}{p_i^{\text{X}} r_i^{\text{X}}} \qquad \text{and} \qquad \text{SNIP}^{\text{Y}} = \frac{3}{m^{\text{Y}}} \sum_{i=1}^{n^{\text{Y}}} \frac{1}{p_i^{\text{Y}} r_i^{\text{Y}}}. \qquad (A3)$$

The SNIP value of the merged journal, referred to as journal XY, then equals

$$\text{SNIP}^{\text{XY}} = \frac{3}{m^{\text{X}} + m^{\text{Y}}} \left( \sum_{i=1}^{n^{\text{X}}} \frac{1}{p_i^{\text{X}} r_i^{\text{X}}} + \sum_{i=1}^{n^{\text{Y}}} \frac{1}{p_i^{\text{Y}} r_i^{\text{Y}}} \right). \qquad (A4)$$

This can be rewritten as

$$\text{SNIP}^{\text{XY}} = \frac{m^{\text{X}}}{m^{\text{X}} + m^{\text{Y}}} \text{SNIP}^{\text{X}} + \frac{m^{\text{Y}}}{m^{\text{X}} + m^{\text{Y}}} \text{SNIP}^{\text{Y}}. \qquad (A5)$$

Eq. (A5) indicates that the SNIP value of journal XY is a weighted average of the SNIP values of journals X and Y. This implies that journal XY has a SNIP value that is in between the SNIP values of journals X and Y.

The above merging property of the revised SNIP indicator illustrates that the indicator behaves in a consistent way. Somewhat informally, the notion of consistency of an indicator, introduced by Waltman et al. (2011) and Waltman and Van Eck (2012a), requires that the ranking of two units relative to each other remains unchanged when both units make the same performance improvement.[14] Based on (A2), the revised SNIP indicator trivially satisfies this requirement.

## Appendix B: Source normalization in a formal mathematical framework

In this appendix, we provide a formal mathematical framework that allows us to show how the revised SNIP indicator corrects for differences in citation practices between scientific fields. The approach that we take is similar to the approach taken in an earlier paper (Waltman & Van Eck, 2010) to study the normalization mechanism of the audience factor of Zitt and Small (2008). A somewhat similar approach is also taken by Zitt (2011) in his analysis of the journal impact factor.

Our approach relies on the following three assumptions:
1. The set of all journals in the database can be partitioned into a number of subsets in such a way that journals in one subset do not cite journals in other subsets. We refer to each subset of journals as a field. Hence, what we assume is the absence of between-field citation traffic.
2. Each year, the number of publications in a field is the same.
3. Each journal has at least one publication that has at least one active reference. In other words, there are no journals that have no active references at all.

It is of course clear that in practice the above assumptions do not hold perfectly well. The results that we present in this appendix therefore do not apply directly in practice. However, depending on the degree to which our assumed idealized world approximates the real world, our results can be expected to hold in an approximate sense. We will get back to the sensitivity of our results to the underlying assumptions at the end of this appendix.

---

[14] The notion of consistency is similar to the notion of independence studied by Marchant (2009a, 2009b) and Bouyssou and Marchant (2011).

We start our mathematical analysis by introducing some notation. For a given year of analysis, we distinguish between citing publications and citing journals on the one hand and cited publications and cited journals on the other hand. Citing publications are publications that appeared in the year of analysis, while cited publications are publications that appeared in the three preceding years. Let $N_1$ and $N_2$ denote, respectively, the number of cited journals and the number of citing journals in a particular field. In practice, the sets of cited and citing journals will often coincide, but our framework does not require this to be the case. Let $m_{k1}$ and $m_{k2}$ denote, respectively, the number of publications of cited journal $k$ and the number of publications of citing journal $k$. We further define

$$M_1 = \sum_{k=1}^{N_1} m_{k1} \tag{B1}$$

and

$$M_2 = \sum_{k=1}^{N_2} m_{k2} . \tag{B2}$$

Hence, $M_1$ and $M_2$ denote the total number of cited and citing publications in a particular field. Notice that assumption 2 above implies that $M_1 = 3M_2$. This follows from the fact that we have three cited years and only one citing year.

Our focus is on the average SNIP value of the cited journals in a field, where each journal is weighted by its number of publications. In other words, we are interested in

$$\mu = \sum_{k=1}^{N_1} \frac{m_{k1}}{M_1} \text{SNIP}_k . \tag{B3}$$

We will say that field differences have been properly corrected for if $\mu$ has the same value in all fields. What we will prove is that in our framework this is indeed the case. Under the three assumptions introduced above, we will show that $\mu$ always equals one.

Based on (A2), we rewrite (B3) as

$$\mu = \frac{3}{M_1} \sum_{k=1}^{N_1} \sum_{i=1}^{n_k} \frac{1}{p_{ki} r_{ki}} , \tag{B4}$$

where $n_k$ denotes the number of citations received by cited journal $k$, $r_{ki}$ denotes the number of active references in the publication from which the $i$th citation to journal $k$ originates, and $p_{ki}$ denotes the proportion of publications with at least one active reference in the journal from which the $i$th citation originates. It follows from assumption 1 above that the citations received by the cited publications in a field coincide with the citations given by the citing publications in the same field. Because of this, (B4) can be reformulated as

$$\mu = \frac{3}{M_1} \sum_{k=1}^{N_2} \sum_{j=1}^{q_k m_{k2}} \sum_{i=1}^{s_{kj}} \frac{1}{q_k s_{kj}} , \tag{B5}$$

where $q_k$ denotes the proportion of publications with at least one active reference in citing journal $k$ and $s_{kj}$ denotes the number of active references in the $j$th publication with at least one active reference in citing journal $k$. Notice that the second summation in (B5) extends over the $q_k m_{k2}$ publications with at least one active reference in citing journal $k$. Assumption 3 above ensures that there is always at least one such publication. Eq. (B5) can be simplified into

$$\mu = \frac{3}{M_1} \sum_{k=1}^{N_2} m_{k2} = \frac{3M_2}{M_1}. \tag{B6}$$

As already noted, assumption 2 above implies that $M_1 = 3M_2$. Hence, combining (B6) with assumption 2, we obtain $\mu = 1$. In other words, we have proven that under the three assumptions of our mathematical framework the average SNIP value of the cited journals in a field always equals one. This means that in our framework the revised SNIP indicator properly corrects for field differences.

The intuition underlying the above mathematical analysis can be summarized as follows. Suppose first that there are no publications without active references. In the case of a publication with $r$ active references, each reference then has a 'value' of $1/r$. This means that the overall value of the active references in a publication equals $r \times (1/r) = 1$. Hence, regardless of the number of active references in a publication, the overall value of the active references always equals one.[15] As a consequence, the overall value of all active references in a field equals the total number $M$ of citing publications in the field. Assuming that citation traffic mainly takes place within the same field and that the publication output in a field is fairly stable over time, there are about $3M$ cited publications that need to share the overall value of $M$ of the active references in a field. The average value received per cited publication then equals $M/3M = 1/3$, which becomes an average value of one after multiplication by a constant value of three. The average SNIP value of the journals in a field equals the average value received per cited publication, and we therefore obtain an average SNIP value of one. This average SNIP value does not depend on the citation practices in a field (e.g., long or short reference lists), and in that sense differences in citation practices between fields have been corrected for.

The above reasoning is no longer valid if there are citing publications that have no active references. The problem is that citing publications without active references do not provide any 'value' to the cited publications in a field, thereby distorting the normalization for field differences. The revised SNIP indicator corrects for this by increasing the value of the active references in citing publications that do have active references. For instance, if half of the citing publications in a journal have no active references, the value of the active references in the other half of the citing publications is doubled.[16] In this way, it is guaranteed that the overall value of all active references in a field equals the total number of citing publications in the field, and this in turn ensures that we always end up with an average SNIP value of one in a field.

The source normalized indicators proposed by Leydesdorff and Opthof (2010), Leydesdorff and Bornmann (2011), and Glänzel et al. (2011) are similar to the revised SNIP indicator but do not correct for citing publications that have no active references. Because of this, these indicators may have a bias against fields in which the share of publications without active references is relatively large. This will typically be fields with a low citation density. Indeed, Leydesdorff and Bornmann as well as Glänzel et al. find empirically that their proposed indicators correct only partially for differences between low citation density fields (e.g., engineering and mathematics) and high citation density fields (e.g., biosciences). Additional empirical evidence is provided by Waltman and Van Eck (2012b).

---

[15] In the literature, this idea of 'fractional citation counting' goes back to Small and Sweeney (1985).
[16] This is taken care of by $p_i$ in (5) and (A2).

Finally, let us briefly comment on the sensitivity of our analysis to the assumptions introduced in the beginning of this appendix. The first assumption is that there is no between-field citation traffic. Violations of this assumption need not be a problem, as long as the citation traffic between two fields takes place in both directions. However, if one field frequently cites another while the reverse is not true (e.g., an applied field citing a more basic field), this will decrease the SNIP values in the former field and increase those in the latter one. The second assumption is that the publication output in a field is stable over time. This assumption is likely to be violated because in many fields there is an increasing trend in publication output (at least in the publication output that is visible in bibliographic databases). In the case of an increasing trend in publication output, the average SNIP value of the journals in a field will be above one. Between-field comparisons of SNIP values may still be valid, provided that publication output increases at the same rate in different fields. The third assumption is that there are no journals without active references. In the case of the revised SNIP indicator, the journals included in the calculation of the indicator are selected in such a way that this assumption is always satisfied (see Section 4).

## Acknowledgment

We would like to thank Ed Noyons and Paul Wouters from the Centre for Science and Technology Studies and Peter Berkvens, Lisa Colledge, M'hamed el Aisati, Michael Habib, Wim Meester, and Henk Moed from Elsevier for their contributions, in many different ways, to the SNIP project.

## References

Bouyssou, D., & Marchant, T. (2011). Bibliometric rankings of journals based on impact factors: An axiomatic approach. *Journal of Informetrics*, *5*(1), 75–86.

Garfield, E. (1979). *Citation indexing: Its theory and application in science, technology, and humanities*. Wiley.

Glänzel, W., Schubert, A., Thijs, B., & Debackere, K. (2011). A priori vs. a posteriori normalisation of citation indicators. The case of journal ranking. *Scientometrics*, *87*(2), 415–424.

González-Pereira, B., Guerrero-Bote, V.P., & Moya-Anegón, F. (2010). A new approach to the metric of journals' scientific prestige: The SJR indicator. *Journal of Informetrics*, *4*(3), 379–391.

Guerrero-Bote, V.P., & Moya-Anegón, F. (2012). A further step forward in measuring journals' scientific prestige: The SJR2 indicator. *Journal of Informetrics*, *6*(4), 674–688.

Leydesdorff, L., & Bornmann, L. (2011). How fractional counting of citations affects the impact factor: Normalization in terms of differences in citation potentials among fields of science. *Journal of the American Society for Information Science and Technology*, *62*(2), 217–229.

Leydesdorff, L., & Opthof, T. (2010). Scopus's source normalized impact per paper (SNIP) versus a journal impact factor based on fractional counting of citations.

*Journal of the American Society for Information Science and Technology*, *61*(11), 2365–2369.

Leydesdorff, L., & Opthof, T. (2011). Scopus' SNIP indicator: Reply to Moed. *Journal of the American Society for Information Science and Technology*, *62*(1), 214–215.

Leydesdorff, L., Zhou, P., & Bornmann, L. (in press). How can journal impact factors be normalized across fields of science? An assessment in terms of percentile ranks and fractional counts. *Journal of the American Society for Information Science and Technology*.

Marchant, T. (2009a). An axiomatic characterization of the ranking based on the *h*-index and some other bibliometric rankings of authors. *Scientometrics*, *80*(2), 327–344.

Marchant, T. (2009b). Score-based bibliometric rankings of authors. *Journal of the American Society for Information Science and Technology*, *60*(6), 1132–1137.

Moed, H.F. (2010). Measuring contextual citation impact of scientific journals. *Journal of Informetrics*, *4*(3), 265–277.

Moed, H.F. (2011). The source normalized impact per paper is a valid and sophisticated indicator of journal citation impact. *Journal of the American Society for Information Science and Technology*, *62*(1), 211–213.

Radicchi, F., & Castellano, C. (2012). Testing the fairness of citation indicators for comparison across scientific domains: The case of fractional citation counts. *Journal of Informetrics*, *6*(1), 121–130.

Small, H., & Sweeney, E. (1985). Clustering the science citation index using co-citations. I. A comparison of methods. *Scientometrics*, *7*(3–6), 391–409.

Waltman, L., & Van Eck, N.J. (2010). The relation between Eigenfactor, audience factor, and influence weight. *Journal of the American Society for Information Science and Technology*, *61*(7), 1476–1486.

Waltman, L., & Van Eck, N.J. (2012a). The inconsistency of the *h*-index. *Journal of the American Society for Information Science and Technology*, *63*(2), 406–415.

Waltman, L., & Van Eck, N.J. (2012b). *Source normalized indicators of citation impact: An overview of different approaches and an empirical comparison*. arXiv:1208.6122.

Waltman, L., Van Eck, N.J., Van Leeuwen, T.N., Visser, M.S., & Van Raan, A.F.J. (2011). Towards a new crown indicator: Some theoretical considerations. *Journal of Informetrics*, *5*(1), 37–47.

Zitt, M. (2010). Citing-side normalization of journal impact: A robust variant of the audience factor. *Journal of Informetrics*, *4*(3), 392–406.

Zitt, M. (2011). Behind citing-side normalization of citations: Some properties of the journal impact factor. *Scientometrics*, *89*(1), 329–344.

Zitt, M., & Small, H. (2008). Modifying the journal impact factor by fractional citation weighting: The audience factor. *Journal of the American Society for Information Science and Technology*, *59*(11), 1856–1860.